\documentclass[prl,aps,10pt,superscriptaddress,onepage]{revtex4}
\usepackage{epsfig,amsmath,amssymb,graphicx,color,calc}
\usepackage{graphicx}
\usepackage{amssymb} \usepackage{epstopdf} \usepackage{booktabs}
\usepackage{rotating} \input{epsf}

 \def\de{\mathrm d}

\let\a=\alpha \let\b=\beta  
   
\let\l=\lambda    
  \let\t=\tau

   \let\r=\rho

\def\ba{\begin{align}} \def\ea{\end{align}}

\def\to{\rightarrow}

\def\de{\mathrm d}

\newcommand{\beq}{\begin{equation}} \newcommand{\eeq}{\end{equation}}

\begin{document}

\title{Quasi-equilibrium in glassy dynamics: a liquid theory approach}

\author{Silvio Franz}
\affiliation{Laboratoire de Physique Th\'eorique et Mod\`eles
  Statistiques, CNRS et Universit\'e Paris-Sud 11,
UMR8626, B\^at. 100, 91405 Orsay Cedex, France}

\author{Giorgio Parisi}
\affiliation{Dipartimento di Fisica,
Sapienza Universit\'a di Roma,
INFN, Sezione di Roma I, IPFC -- CNR,
P.le A. Moro 2, I-00185 Roma, Italy
}

\author{Pierfrancesco Urbani}
\affiliation{IPhT, CEA/DSM-CNRS/URA 2306, CEA Saclay, F-91191 Gif-sur-Yvette Cedex, France}

\begin{abstract}
  We introduce a quasi-equilibrium formalism in the theory of 
  liquids in order to obtain a set of coarse grained 
  dynamical equations for the description of long time glassy relaxation.
  Our scheme allows to use typical approximations devised
  for equilibrium to study glassy dynamics. 
After introducing dynamical Ornstein-Zernike relations,
we focus on the
  Hypernetted Chain (HNC) approximation and a recent closure scheme  developed by
  Szamel. In both cases we get dynamical equations that have the
  structure of the Mode-Coupling Theory (MCT) equations in the long time
  region. The HNC approach, that was so far used to get equilibrium
  quantities is thus generalized to a fully consistent scheme where
  long-time dynamic quantities can also be computed. In the context of
  this approximation we get an asymptotic description of both
  equilibrium glassy dynamics at high temperature and of aging dynamics
  at low temperature. The Szamel approximation on the other hand is
  shown to lead to the canonical MCT equations obtained by G\"otze for
  equilibrium dynamics.  We clarify the way phase space is sampled
  according to MCT during dynamical relaxation.

\end{abstract}
\maketitle
%DOVREMMO VEDERE, ED AGGIUNGERE NEL TESTO, CHE LE EQ.NI DINAMICHE HNC
%POSSONO ESSERE OTTENUTE COL FORMALISMO SU-SY DALLA VERA DINAMICA
%-NEWTON O LANGEVIN-, TRASCURANDO SISTEMATICAMENTE I DIAGRAMMI CHE
%VENGONO DALLA CONSERVAZIONE DELLA MASSA. QUESTO DOVREBBE ESSERE QUASI
%TAUTOLIGICO. 
The relation between dynamical slowing down and
equilibrium free-energy landscape is a longly debated issue in the theory of glassy relaxation of low
temperature liquids  \cite{BB11, BBBCS11}. 
Free energy landscape \cite{Go69, SW82, BB09} and pure dynamical approaches \cite{RS03, BBBCS11} provide alternative pictures of   %Vedi libro Luca.
the glass transition \cite{BG13}. 
%Moreover schematic solvable mean field models where 	%% Opposizione KCM and Landscape approaches - RFOT 
%%where both equilibrium and dynamics can be studied exactly, 
%dynamical
%arrest and ergodicity breaking appear as different sides
%of the same coin \cite{KTW89, Cavagna-Spin-Glasses}, can be sided by Kinetically Constrained Models \cite{KCM} where the glass transition is purely dynamical.  
%This observation is at the basis of the ``Random First Order Transition Theory'' of the glass transition.      %%Rimosso dalla version 1
%In this letter we investigate the relation between dynamics and equilibrium within first principle liquid theory.
At the theoretical level,
%beyond these schematic models, 
the standard techniques and approximations involved in the study of
equilibrium and dynamics of realistic liquid models are considerably
different. {
Here we overcome this difficulty relating the dynamical properties of interest with the ones of an appropriate auxiliary system with Boltzmann-Gibbs distribution.}
%resulting in quantitative disagreements in observables computed in both frameworks.
%even at the
%mean field level. 
%At the dynamical level 									%%Rimosso dalla versione 1 e sostituito con In dynamics

In dynamics, 											%%Aggiunto
Mode Coupling Theory                               		
(MCT), based on projection operator techniques, provides an approximation for an exact evolution equation   
of the density-density correlator and requires the static structure		 %% 
function as an input \cite{Go09}. In equilibrium, the replica
method is used in conjunction with approximations like the Hypernetted Chain
(HNC)
 %partial resummation of virial series						%%Rimosso dalla versione 1
 \cite{MP96} or the small
cage expansion \cite{PZ10}. Both approaches differ on specific
predictions, but agree on the same general picture. 
%Ergodicity
%breaking appears to be driven by critical fluctuations (dynamical
%heterogeneities) with the same universal characteristics and growth of			%%Rimosso dalla version 1
%correlations.  
In recent times there has been considerable effort to
reconcile the
two 																%%Aggiunto all a version 1
approaches in the ``$\beta$ regime'', where, after a
first fast relaxation, correlation functions remain close to a plateau
value.  
%On the one hand, 												%%Rimosso dalla version 1
Szamel \cite{Sz10} has derived the MCT
equation for the non-ergodicity parameter using a particular closure
of the replicated YBG hierarchy for the density correlation functions.
%On the other, an important dynamical quantity, 						%%Rimosso
Moreover 														%%Aggiunto
the celebrated exponent						
parameter of MCT, has been related to amplitude ratios between
equilibrium correlation functions \cite{CFLPRR12} and has been
computed in the framework of replicated liquid theory \cite{FJPUZ12,
  FJPUZ12Long, KPUZ13}.  
%  Unfortunately, general approximation frameworks
%capable to describe in a unified way both equilibrium and dynamics are
%at present lacking.  
One of the difficulties in unifying equilibrium and dynamical approaches lies in the necessity of
including conservation laws (mass, momentum and energy) in the
dynamical description at short times.  However this problem should
disappear in the long time ``$\alpha$ regime''
where correlations decay below their plateau value.
% and where									%%Rimosso 
%In that case quasi-stationarity should hold 
%in the dynamical equations.
%In this paper 													%%Rimosso
Here															%%Aggiunto
we show that progress can be achieved using a coarse
grained description where the details of the short times are neglected
and one concentrates on the slow part of the dynamics.  
%Despite the fact that glasses are out of equilibrium, 					%%Rimosso
Emerging properties of
glassy relaxation, as temperature-time superposition principle
\cite{Go09} or the appearance of effective temperatures  \cite{CKP97}, 
can be rationalized through the hypothesis that
quasi-equilibrium principles drive the dynamics during the phase space
exploration in the long time regime \cite{FV00}.  
In a nutshell, 													%%Rimosso
the two steps glassy relaxation at low temperature can be described as a
first rapid decay inside a metastable state, 
followed by a very slow relaxation in a complex landscape of
metastable states.
%in the long time $\alpha$ region. 									%%Rimosso
%The short time
%regime is sensitive to the details of the dynamics 
%dynamical rules of evolution, 									%%Rimosso
%while the long time one is expected to be independent from
%such details and it 
This relaxation can be thought as a random walk in phase space
where the states that are available at a given time are selected
according to their Boltzmann weight.
%one can
%describe slow relaxation in the $\alpha$ region as a random walk in a complex landscape of metastable metastable states,
%where the states that are available at a given time are selected
%according to their Boltzmann weight. 
This insight has been formalized in 
%a recent paper 												%%Rimosso
\cite{FP12} through
the introduction of an appropriate Markov chain construction
\cite{KK07,KZ11}. Consider
a system in a discrete time dependent configurations specified by $S_\tau$ and
subjected to a Hamiltonian $H({S})$ and an overlap function
$Q(S,S')=\int \de x\de y\, w(x-y) \rho_S(x)\rho_{S'}(y)$, that					%Citare Giorgio
measures the similarity between density profiles of couples of
configurations \footnote{The function $w$ doesn't affect
our results, provided it is short ranged.}.  
One can define the {\it Boltzmann pseudodynamics} 
with the following discrete time Markov Chain
\begin{eqnarray}
M(S_{\tau+1}|S_\tau)=\frac{e^{-\beta H(S_{\tau+1})}}{Z[S_\tau]}
\delta\left(Q(S_{\tau+1},S_\tau)-{ C}(\tau+1,\tau)\right)\label{chain}
\end{eqnarray}
which implements the previous idea, assuming that the states available
at a given time $\tau+1$ are the ones that lie at an overlap
$C(\tau+1,\tau)$ from the configuration at time $\tau$.  This construction has been employed in mean field spin
glasses where Boltzmann pseudodynamics provides an exact coarse grained
version of slow dynamics when an infinitely long chain is considered
and the value of $C(\tau+1,\tau)$ is such that the constraints do not
affect the free-energy of the system \cite{FP12}. This result is
remarkable at the fundamental and practical levels. 
%At the fundamental level							%%Rimosso 
It provides a basic probabilistic description of how
configuration space is explored during glassy relaxation. 
%At the practical level,								%%Rimosso 
Moreover											%%Aggiunto
it opens the possibility to study long time dynamics
through equilibrium techniques.
%The scope of this letter is							%%Rimosso
We want											%%Aggiunto
to study Boltzmann pseudodynamics in
glassy liquids, and to discuss its relation with present dynamical
theories, in particular, with the Mode Coupling Theory (MCT) of the
glass transition and the theory of aging dynamics as is known from
schematic spin models.  
%As usual, 										%%Rimosso
We will need approximations to deal
with the strongly interacting character of the systems.  We will
concentrate in 
%liquid theory approximation leading to closed equations	%%Rimosso
%for density-density correlation functions: 
1) the well known
Hypernetted Chain approximation 2) the Szamel closure scheme of the
replicated YBG hierarchy \cite{Sz10}.  Our main results are:		%%Rimosso 
%the following:
\begin{itemize}
\item We introduce a long time dynamical version of the Ornstein-Zernike (OZ) equations. 
This equations can be use within any selected closure scheme and are quite general MCT-like equations.
%Within the HNC approximation we get fully self-consistent
%  dynamical equations for the correlation and response functions that 
%%  consistently											%%Rimosso 
%  describe the evolution of liquids at low temperature. 
\item We use the HNC approximation in order to close the OZ dynamical equation.
We obtain a complete set of new dynamical equations that describe the evolution of liquids at low temperature.
Moreover,   starting from an equilibrium initial condition our equations
  reduce to a single one for the correlation function. This equation
  displays the universal features of MCT with a singularity at a previously identified dynamical
  transition point \cite{MP96}. 
  %In particular, 										%%Rimosso
  Relaxation in the
  early and late $\beta$ regime is characterized by power laws whose
  {respective} exponents $a$ and $b$ verify the MC relation $\Gamma (1-a)^2
  /\Gamma(1-2a) =\Gamma(1+b)^2 /\Gamma(1+2b)=\lambda$. The exponent
  parameter $\lambda$ coincides with the one computed in \cite{FJPUZ12, FJPUZ12Long}. 
  Differently from MCT which needs the external input of
  the static structure factor, we get here a complete set of equations
  determining static and long time dynamical quantities.
\item Always within HNC, starting from a strongly off-equilibrium situation, we find an
  aging regime. Modulo the identification of a soft mode of slow
  relaxation, the gross picture follows the main lines found in mean
  field spin glasses \cite{CK93}. The system falls asymptotically in a
  non-equilibrium state where time translation invariance and
  fluctuation dissipation theorem are violated.  
  %%We compute										%%Rimosso
  %self-consistently the emerging effective temperature.			
\item 
%As an alternative approximation, we use a scheme recently		%%Rimosso
%  proposed by Szamel.  We show that this approximation 
  We use Szamel's scheme that								%%Aggiunto
  allows to
  derive the canonical Mode Coupling equation for the slow part of the
  momentum dependent correlator in the critical regime. We thus unveil
  that, 
  %as described by 										%%Rimosso
  within												%%Aggiunto
  MCT, the relaxation in the whole $\alpha$
  regime consists in quasi-equilibrium exploration of the
  configuration space that explains the insensitivity of this regime
  from the details of 
  %dynamical evolution									%%Rimosso
  dynamics											%%Aggiunto
  \cite{gotze1991liquids,szamel1991mode}.
\end{itemize}

{\it Boltzmann pseudodynamics.} 
In order to study the pseudodynamic process one can start 
from the probability of a trajectory {of length $L$}
\begin{gather}\label{probtraj}
P(S_L,S_{L-1},...,S_0)=\frac{e^{-\beta_0 H[S_0]}}{Z(\beta)}\cdot\\
\cdot \prod_{\tau=0}^{L-1}  
\frac{e^{-\beta H[S_{\tau+1}]}}{Z[S_\tau,C(\tau+1,\tau)]}\delta\left(Q(S_{\tau+1},S_\tau)-C
(\tau+1,\tau)\right)\nonumber
\end{gather}
where $Z(\beta)=\sum_{S}e^{-\beta H[S]}$ and $Z[S,\,C]=\sum_{S'}e^{-\beta H[S']}\delta\left(Q(S',S)-C
\right)$ are normalization factors.
The choice of $\beta_0=\beta$ corresponds to choose an equilibrium
initial condition at temperature $\beta$, while $\beta_0=0$ is useful
to model a high temperature quench and an aging situation. The
probability (\ref{probtraj}) can be used to compute averages of
observables or the free-energy of the system as a function of time. 
%To use it in practice,		%%Rimosso 
One way to deal with this construction 
is {to use} the replica method
%that reduces to the substitution of
and substitute
the partition functions $Z[S_\tau,\,C(\tau+1,\tau)]$ in the denominator
of (\ref{probtraj}) with 
${Z^{-1}[S_\tau,\,C(\tau+1,\tau)]}\to Z[S_\tau,\,C(\tau+1,\tau)]^{n_\tau-1}$. 
As usual, the numbers
of replicas $n_\tau$, that here can depend on time, is kept integer for
intermediate computations and finally sent to zero. 
%For each time
%$u$ there is a master replica, $S_1(u)$, and $n_u-1$ slave replicas
%$S_a(u)$ $a=2,...,n_u$. Both master and slave replicas at a given time
%$t$ are coupled to the master replica at the previous time $t-1$
%through the constraint. 
Enforcing the constraint in (\ref{chain}) through a Lagrange
multiplier $\nu(\tau)$, and expliciting the partition functions as
sums over configurations, 
we obtain formally the expressions of the equilibrium of a system made of a mixture of different
particle spices so that 
%(one for each time and replica index) 		%%Rimosso
static approximation schemes
can be
%streightforwardly
applied. 
{Each system is indexed by a time label $\t$ that
encodes for its position along the chain and a replica index $a$ that goes from 1 to $n_\t$. We define for notational simplicity
Greek indexes that code 
for both time and replica indexes, e.g. $\alpha\equiv (\tau,a)$. }

{\it Dynamical Ornstein-Zernike equations.}  Let us introduce the
basic fields that we 
%want to 					%%Rimosso
study.  
%%Suppose that			%%Rimosso 
We start with a
chain of systems where each system is composed by $N$ particles interacting 
through a pair potential $\phi(x)$. 
% \begin{gather}
% \rho_a(x)=\sum_{i=1}^N \delta\left(x-x_i^{(a)}\right)\\
% \rho_{ab}(x,y)=\rho_a(x)\rho_b(y)-\rho_a(x)\delta_{ab}\delta(x-y)\:.
% \end{gather}

%SF
 
%%can						%%Rimosso 
The replicated one and two points density functions are 
\begin{gather}
\rho_\alpha(x)=\sum_{i=1}^N \delta\left(x-x_i^{(\alpha)}\right)\\
\rho_{\alpha\beta}(x,y)=\langle\rho_\alpha(x)\rho_\beta(y)\rangle-\langle\rho_\alpha(x)\rangle\delta_{\alpha,\beta}\delta(x-y)\:.
\end{gather}
Our starting point 
%of our analysis 			%%Rimosso
are the OZ equations for a mixture of particles \cite{GS93, Gi92}

%\begin{gather}
%\ln[h_{\alpha\beta}(x,y)+1]+\beta\phi_{\alpha\beta}(x,y)=h_{\alpha\beta}(x,y)-c_{\alpha\beta}(x,y)
%\end{gather}
%where $\phi_{\alpha\beta}(x,y)\equiv \phi_{a\tau;b\sigma}(x,y)
%=\delta_{\tau,\sigma}\delta_{a,b}\phi(x-y)+\delta_{b,1}\delta_{\tau,\sigma+1}w(x-y)
%\nu(\sigma)$ 
%contains the inter particle potential at equal time and replica indexes
%%%as well as   %%Rimosso
%and 		%%Aggiunto
%the Lagrange multiplier constraining the value of the overlap at consecutive times.   

\beq\label{OZrep}
c_{\alpha\beta}(x,y)=h_{\alpha\beta}(x,y)-\rho \sum_\gamma
%_{c=1}^{n}
\int \de zh_{\alpha\gamma}(x,z)c_{\gamma\beta}(z,y)\:, 
\eeq 
where we
have supposed that the density $\langle \rho_\alpha(x)\rangle =\rho$
is constant. 
Here $\rho^2h_{\alpha\beta}(x,y)=\rho_{\alpha\beta}(x,y)-\rho^2$, while $c_{\alpha\beta}(x,y)$ defines 
the direct correlation function.

%Notice that for $\nu(t)\equiv 0$, these equations are
%formally identical to the usual HNC equations for replicated liquids
%and admit 
%%%well known 
%time independent solutions that describe a glass
%transition at well defined 
%%%temperature $T_d$   %%Rimosso
%dynamical point             %%Aggiunto 
%in the static 
%%replica			%%Rimosso
%formalism \cite{MP96}.
% in the $\beta$ region.
We want to use this equation to study slow dynamics. As explained in
\cite{FP12} this can be done seeking for
continuous solutions in the infinite chain limit, where the time
variables $t=\tau/L$ etc. become continuous and $\nu(t)\to 0$ for all $t$.
% To do this we will search a \emph{causal} replica
% symmetric solution for that equations that encodes the Boltzmann
% pseudodynamic ansatz. The key point is that the constraints in
% (\ref{chain}) force the chain of systems to break the replica
% symmetry.  
%The consequence is that the most simple replica ansatz is
%described by
%The simplest parametrization allowing us to achieve this goal, writing 		%%Rimosso
%explicitly time and replica indexes, reads:
As usual in the replica method, we need to choose a parametrization of the
replica matrix $h_{\alpha\beta}(x,y)$ that allows the analytic continuation to $n_t\to 0$. In \cite{FP12} an appropriate form to describe long time dynamics was shown to be
 \beq\label{parametrization}
\begin{split}
h_{\alpha\beta}(x,y)&=h(x,y;s,u)+\delta_{ab}\delta (s-u)\Delta h(x,y;s,u)\\
&+\Theta(u-s)\delta_{a1}\Delta h(x,y;u,s)\\
&+\Theta(s-u)\delta_{b1}\Delta h(x,y;s,u)
\end{split}
\eeq 
where 
%here				%%Rimosso 
$\a=(s,a)$ and $\b=(u,b)$.
%$s$ and $u$ are time indexes, \textcolor{red}{while $a$ and $b$ are
% replica indexes that select respectively a specific replica of the systems at the position $s$ and $u$.}
The Heaviside function $\Theta(x)$ is here defined equal to 1 when $x$ is strictly positive and zero
otherwise.
The ansatz (\ref{parametrization}) is consistent with causality and respects the replica symmetry of the problem. It can be shown that 
this is equivalent to assume the self-averaging  of space averaged correlation  and response functions 
with respect to the thermal noise.
% Note that the expression (\ref{parametrization}) is just a
%parametrization. 
					%%Aggiunto
%the direct correlation function. 			%%Rimosso
%  Using 
% %At this point we can put the parametrization
% (\ref{parametrization}) into the HNC and Ornstein-Zernike equations to
% find the actual solution of the equations.
In the limit 
%%of zero replicas 				%%Rimosso
$\{n_u\}\to 0$, $h(x,y;s,u)$ represents the 
normalized density-density space and time dependent correlation function 
along the chain; remarkably, 
in the long chain limit $\de u\simeq 1/L$ and the function
\beq 
R_h(x,y;s,u)=\beta \Theta(s-u)\frac{\Delta h(x,y;s,u)}{\de
u}
\label{ar} 
\eeq
for $s>u$ appears to be the response 
function that encodes how the density in $x$ at time $s$
varies when a small
% perturbation
%SF
perturbing pressure is applied to the system in $y$ at time $u$.  
Analogous expressions hold for 
$c_{\alpha\beta}(x,y)$ that therefore defines not only a direct time dependent correlation function but also a direct response function.
The equal time quantities $\Delta h(q;s,s)$ and $\Delta c(q;s,s)$ 
%are
%actually $s$-independent and 
%%represent the jump from the instantaneous
%%to the plateau values of the correlation functions as effect of the			%%Rimosso
%%short time dynamics that we neglect.  
encode the information on the short time dynamics, {namely 
the jump in correlation from the 
instantaneous to the plateau value.}	
%As
%the effect of short time dynamics is condensed in a single step of the
%chain, we can expect that $h(x,y;s,u)$ is
%discontinuous at equal time with $\Delta h (x,y;s,s)=h(x,y,s,s)-\lim_{u\to
% s}h(x,y;s,u)$ representing the jump in correlation from the 
%instantaneous to the plateau value.
%In what follows
%We will search for a space translational			%%Rimosso
%invariant solution of the equations.  
Assuming space translational invariance,			%%Aggiunto
%By										%%Rimosso
using (\ref{ar})
and 
%by										%%Rimosso 
taking the infinite chain limit, the OZ equations
%\textcolor{red}{in doing the matrix product in (\ref{OZrep})}, we can derive from the
%-Zernike equation in Fourier space the dynamical equations
(\ref{OZrep}) in Fourier space become
\begin{gather}
\label{correlation}
\begin{split}
0&=W_q[h,c]+\frac{\r}{\beta}\int_0^u \de z\, h(q;s,z)R_c(q;u,z)+\\
&+\frac{\r}{\beta}\int_0^s\de z\, R_h(q;s,z) c(q;z,u)
\end{split} \\
W^{(R)}_q[R_h,R_c]=-\frac{\r}{\beta}\int_s^u \de z R_h(q;z,s)R_c(q;u,z)\label{response}\\
\Delta h(q;s,s)=\Delta c(q;s,s)+\r \Delta h(q;s,s)\Delta c(q;s,s)\label{diagonalOZ}
\end{gather}
where we have defined 
$W_q[h,c]=c(q;s,u)-h(q;s,u)+\r[h(q;s,0)c(q;u,0)+h(q;s,u)\Delta c(q;u,u)+c(q;s,u)\Delta h(q;s,s)]$
and 
$W^{(R)}_q[R_h,R_c]=R_c(q;u,s)-R_h(q;u,s)+\r R_h(q;u,s)\Delta c(q;u,u)+\r R_c(q;u,s)\Delta h(q;s,s)$.
These equations are valid for $u\geq s$; the equivalent
ones hold for $s>u$. 
We notice that both (\ref{correlation}) and (\ref{response}) are compatible with equilibrium where time translational invariance (TTI) and fluctuation dissipation theorem $R_{h}(q,s-u)=-\beta \de h(q, s-u)/\de s$ hold. In this case they reduce to the single equation
\beq
F_q[h]=\int_0^s\de z \frac{\de h(q;z)}{\de
  z}\left[c(q;s-z)-c(q;s)\right]
\label{OZ-TTI}
\eeq where $F_q[h]=c(q;s)-h(q;s)+\r[h(q;s)\Delta c(q)+c(q;s)\Delta
h(q)+c(q;0)h(q;s)-(h(q;s)-h(q;0))c(q;s)]$.  
This equation has the structure  the long time MCT
equation  \cite{Go09} where the time derivative is neglected and the 
memory kernel is represented by the direct correlation function.
%This equation has the structure of a long time 
%MCT equation whose memory kernel is represented by the
%direct correlation function.
%The equation above is deduced only 
As in the usual static case, the OZ equations are useful in the
context of closure approximations that provide a second equation
that relates the response and correlation functions.
{ The final
equations depend on the closure one uses. We now proceed to the analysis of 
the HNC approximation and the Szamel closure. 
}

{\it The HNC closure: equilibrium and aging solution.}
We study now the closure provided by the HNC approximation for 
a particle mixture \cite{Hansen}. This closure has proved to provide a coherent qualitative picture of glassy phenomena on which we focus, despite it fails to give reliable quantitative results. In the present case it can be rewritten as
\begin{gather}\label{HNCclosure}
\begin{split}
\ln[h_{\alpha\beta}(x,y) &+ 1]+\beta\phi_{\alpha\beta}(x,y)=\\
&=h_{\alpha\beta}(x,y)-c_{\alpha\beta}(x,y)
\end{split}
\end{gather}
where for us $\phi_{\alpha\beta}(x,y)\equiv \phi_{a\tau;b\sigma}(x,y)
=\delta_{\tau,\sigma}\delta_{a,b}\phi(x-y)+\delta_{b,1}\delta_{\tau,\sigma+1}w(x-y)
\nu(\sigma)$ 
contains the inter particle potential at equal time and replica indexes
%%as well as   %%Rimosso
and 		%%Aggiunto
the Lagrange multiplier constraining the value of the overlap at consecutive times.   
Plugging the parametrization (\ref{parametrization}) into (\ref{HNCclosure}) we obtain for $s\ne u$
\begin{gather}
\label{hR}
\ln[h(x;s,u)+1]=h(x;s,u)-c(x;s,u)\\
R_c(x;s,u)=R_h(x;s,u)\frac{h(x;s,u)}{h(x;s,u)+1}\:.\label{hR2}
\end{gather}
%Notice				%%Rimosso 				%%Aggiunto
Together with (\ref{diagonalOZ}) we obtain that the quantities $\Delta h(q;s,s)\equiv \Delta h(q)$ and $\Delta c(q;s,s)\equiv \Delta c(q)$ are $s$ independent.
The dynamical equations
(\ref{correlation}-\ref{hR2}) provide a complete set of equations that
can 
%in principle 					%%Rimosso
be solved in time.  
%Moreover, 					%%Rimosso
They are
reparametrization invariant \cite{CK93, Cu03}: knowing a solution
$h(q;s,u)$, $R_h(q;s,u)$, a whole family is obtained by putting
$h'(q,s,u)=h(q,f(s),f(u))$ and $R_h'(q,s,u)=\frac{\de f(u)}{\de
  u}R_h(q;f(s),f(u))$ (and the analogous expressions for $c$ and
$R_c$) provided that $f(x)$ is an increasing function 
%such that		%%Rimosso
with				%%Aggiunto 
$f(0)=0$.
% <$R_h'(q,s,u)=\frac{\de f(u)}{\de u}R_h(q;f(s),f(u)$.
% For the interesting case $\nu(t)\to 0$ the equation become
% reparameterization invariant.  If we know a solution for them, another
% solution can be obtained by choosing an increasingly monotonous
% function $\phi(t)$ and by putting 
%% At this point we are equipped to study the equilibrium dynamics, where		%%Rimosso 
 Now we want to study the equilibrium dynamics.
 %For generic temperatures, 
%%solutions with 			%%Rimosso
%non-trivial 
%solutions					%%Aggiunto
%%time dependence, 			%%Rimosso
%require non zero 
%%%value of				%%Rimosso
%$\nu(t)$, implying a change in the free-energy. 
%%of the system. 			%%Rimosso
%However, 
{For $T\to T_d^+$ a nontrivial solution 
emerges that satisfies equilibrium, (\ref{hR}) becomes TTI and trough FDT (\ref{hR2}) reduces to the derivative of (\ref{hR})}
% This means that we
% are choosing a particular "gauge" in the space of solutions.
% Clearly this
% is equivalent to choose a gauge in the space of the solutions. 
% defined by
%\begin{gather}
%\ln[h(x;s)+1]=h(x;s)-c(x;s)
%\end{gather}
The phenomenology displayed by the resulting 
%dynamical				%%Rimosso 
equation is similar
to the one of conventional MCT and its solution can be identified with
the scaling function determining the correlation function within the
temperature-time superposition principle \cite{Go09}.   Equation (\ref{OZ-TTI}) is invariant
under time rescaling $t\to A t$ for positive $A$, and the condition
for the existence of a decaying solution is that the operator $\frac{\de
F_q[h]}{\de h(k)}$ has a zero mode ${\cal K}_0(q)$. By using this fact  we can derive an analytic expression for the MCT
exponent parameter \footnote{The detailed derivation will be given elsewhere.}
%(see Supplementary information) 
\beq\label{lambda}
\l=\frac{\int \de^D x {\cal K}_0^3(x)/(1+h(x))^2}{2\r\int \frac{\de^D q}{(2\pi)^D}{\cal K}_0^3(q)[1-\r\Delta c(q)]^3}\,. 
\eeq 
Remarkably, this expression coincides with the one computed in \cite{FJPUZ12, FJPUZ12Long}
using a the static HNC approach.  
%We stress
%that our result is obtained using a full dynamical description.
%For		%%Aggiunto
%the $\alpha$ relaxation time 
%%$\tau_\alpha$ 		%%Rimosso
%%can also be extracted from the equation, through standard MCT analysis and one find   %%Rimosso
%one finds the standard MCT relations					%%Aggiunto
% $\tau_\alpha \sim |T-T_d|^{-\gamma}$ with $\gamma=\frac{1}{2a}+\frac{1}{2b}$. 
%
Below the dynamical glass transition, the dynamical equations 
(\ref{correlation}) and (\ref{response})
under the assumption of a high temperature 
initial condition and loss of memory of the initial condition 
$h(q;s,0)=c(q;s,0)=0$ admit a solution that describes an aging regime. 
%   However in this case the term $W_q[c,h]$ in the equation (\ref{correlation}) 
% have to be replaced by $W_q[c,h]-\r h(q;s,0)c(q;u,0)$ in order to discard the equilibrium initial condition.
The resulting equations could be solved in principle using the modified FDT relation 
$R_h(q;t,s)=X\beta \partial h(q;t,s)/\partial s$ \cite{CK93}.  
%As usual,			%%Rimosso 
The
combination $T/X$ is interpreted as an effective temperature
\cite{CKP97}, and 
%coherently			%%Rimosso 
$X$ is independent of $q$.  As in \cite{CK93}, the FDT ratio $X\in [0,1]$ is fixed by the
requirement that the equation for the response function in the limit
$u\to s$ admits a non vanishing solution. 
%Again 			%%Rimosso
This is only possible in presence of a zero mode and it turns out that
this is equivalent to the marginal stability condition in spin glass
dynamics \cite{CK93}.
The actual computation of X as a function of the control parameters turns out to be numerically quite invoved and we leave it for future works.

{\it A different closure: conventional MCT equations.}  The crucial
point to obtain the dynamical equations
(\ref{correlation}-\ref{OZ-TTI}) 
%has been 		%%Rimosso
is				%%Aggiunto
that, modulo a different
interpretation of the replica indexes, there is formal coincidence of
%replica 			%%Rimosso
pseudodynamic OZ equations with the ordinary equations of
liquid mixtures. This coincidence, which can be traced in the
symmetry of the pseudodynamic effective Hamiltonian under replica index exchange, should be
respected by any consistent approximation scheme. 
%describing the correlation functions. 			%%Rimosso
It has been 
%recently 			%%Rimosso
observed in \cite{Sz10}
that the MCT equation specifying the non-ergodicity parameter can be derived
from replicated OZ equation, through a defined closure
of Yvonne-Born-Green (YBG) hierarchy leading to the following
approximation for the non diagonal elements of the replica directed
correlation function $c_{\alpha\beta}(q)$ for $\alpha\ne \beta$
\begin{eqnarray}
c_{\alpha\beta}(k)=\int dq \; V(k,q) h_{\alpha\beta}(q)
h_{\alpha\beta}(k-q)
\label{sz}
\end{eqnarray}
where the  $V(k,q)$ is the Mode Coupling vertex
given in terms of the static direct correlation function $c_0(q)$ as
\begin{eqnarray}
V(k,q)=\frac{1}{16\pi^3 k^2} [{\hat {\bf k}}\cdot ({\bf q}
c_0(q)+({\bf k-q})c_0(k-q))]^2,  
\end{eqnarray}
which is independent of the replica indexes.  
%Consistently with our previous comment, 		%%Rimosso
We can then interpret (\ref{sz}) in pseudodynamics
with $\alpha\to (a=1,t)$, $\beta\to (a=1,s=0)$ with $t>0$. Assuming
TTI, {we get the expression fo the dynamical 
direct correlation function
\begin{eqnarray}
  \label{eq:direct}
c(k,t)=\int dq \; V(k,q) h(q,t)h(k-q,t).   
\end{eqnarray}
}
Inserting {the previous equation} into (\ref{OZ-TTI}),
after some simple algebra we obtain exactly the G\"otze mode coupling
equation for the slow part of the relaxation.  
From these equations it can be derived the expression for the exponent parameter whose expression coincides with the one derived in \cite{Ri13}.
We notice that in
\cite{La00} MCT has been generalized to describe aging below $T_d$. We
leave for future work the analysis of possible generalizations Szamel
closure to this case. \\
{\it Conclusions.}
%Summarizing, 					%%Rimosso
We have shown that within liquid theory approximations, under marginal
conditions, Boltzmann pseudynamics gives rise to a dynamical picture
strictly following the MCT phenomenology. We have been able to derive
dynamical MCT-like equations in a coherent scheme that includes both
equilibrium and aging dynamics in the context of HNC approximation.
These equations allow to describe equilibrium dynamics on approaching
$T_d$ and aging dynamics below $T_d$. We have computed the zero mode
and the $\lambda$ parameter from these equations. In equilibrium both
these quantities coincide with the ones computed with completely
static techniques in \cite{FJPUZ12}. In aging, the description is
coherent with the $p$-spin picture \cite{CK93}, but goes beyond this
schematic model.
%In this case we have derived an expression for the FDT ratio. 		%%Rimosso
We also found that the conventional
%G\"otze			%%Rimosso 
MCT equation can be obtained from our replica-dynamic Ornstein-Zernike
equations generalizing to pseudo-dynamics a closure scheme of YBG
hierarchy proposed by Szamel. Our analysis demonstrates that critical
MC and MC-like dynamical behavior in equilibrium and aging is
equivalent to quasi-equilibrium sampling of configuration space. We
expect a similar picture to hold in other approximation schemes. In
particular it will be interesting to study the case of small cage
approximation that has been found to give exact result in the limit of
high dimensionality \cite{KPZ12}. These are mean field like results
and the usual caveats apply.
%\textcolor{blue}{In real systems
%the sharp dynamical transition is replaced by a dynamical cross-over 
%due to activated processes.}
{It is well known that due to activated processes the
  sharp transition becomes a dynamical cross-over in real life.}
%is spurious. 
{One should then address the question if our Boltzmann
  pseudoynamics description is valid beyond mean field.  To our view,
  quasi-equilibrium exploration of configuration space is a
  fundamental property of glassy dynamics that should also hold when
  mean field theory fails. Further work will be needed to test this
  conjecture.}  {Moreover it could be interesting to
  see what are the prediction of this approach to the theory of
  dynamical fluctuations in the $\alpha$ regime. Recently it has been
  shown in \cite{FPRR11} using equilibrium techniques that the
  dynamical fluctuations in the $\b$ regime are described by a scalar
  cubic field theory in a random field.  It could be very interesting
  to see if our approach can be used to see if these results extends
  to the $\a$ regime.}

\begin{center}
 {\bf Acknowledgments} 
\end{center}

We thank G. Szamel, F. Ricci-Tersenghi, S. Torquato and F. Zamponi for discussions and S. Sastry for careful reading of the manuscript. 
SF thanks the Physics Department of the Rome university ``La Sapienza'' for
hospitality.  The research of SF has received funding from the
European Union, Seventh Framework Programme FP7-ICT-2009-C under grant
agreement n. 265496. The European Research Council has
provided financial support through European Research Council Grant
247328.
P.U. acknowledges the Physics Department of University of Rome La Sapienza and the 
LPTMS of the University of Paris-Sud where most of this work has been done. 
He acknowledges also the support of the ERC grant NPRGGLASS.

%\cite{BB09}

%\cite{Ca09}

%\bibliography{BPD}{}

\begin{thebibliography}{31}
\expandafter\ifx\csname natexlab\endcsname\relax\def\natexlab#1{#1}\fi
\expandafter\ifx\csname bibnamefont\endcsname\relax
  \def\bibnamefont#1{#1}\fi
\expandafter\ifx\csname bibfnamefont\endcsname\relax
  \def\bibfnamefont#1{#1}\fi
\expandafter\ifx\csname citenamefont\endcsname\relax
  \def\citenamefont#1{#1}\fi
\expandafter\ifx\csname url\endcsname\relax
  \def\url#1{\texttt{#1}}\fi
\expandafter\ifx\csname urlprefix\endcsname\relax\def\urlprefix{URL }\fi
\providecommand{\bibinfo}[2]{#2}
\providecommand{\eprint}[2][]{\url{#2}}

\bibitem[{\citenamefont{Berthier and Biroli}(2011)}]{BB11}
\bibinfo{author}{\bibfnamefont{L.}~\bibnamefont{Berthier}} \bibnamefont{and}
  \bibinfo{author}{\bibfnamefont{G.}~\bibnamefont{Biroli}},
  \bibinfo{journal}{Rev. Mod. Phys.} \textbf{\bibinfo{volume}{83}},
  \bibinfo{pages}{587} (\bibinfo{year}{2011}),
  \urlprefix\url{http://link.aps.org/doi/10.1103/RevModPhys.83.587}.

\bibitem[{\citenamefont{{For a review, see the contributions}}(2011)}]{BBBCS11}
\bibinfo{author}{\bibnamefont{{For a review, see the contributions}}}, in
  \emph{\bibinfo{booktitle}{Dynamical Heterogeneities and Glasses}}, edited by
  \bibinfo{editor}{\bibfnamefont{L.}~\bibnamefont{Berthier}},
  \bibinfo{editor}{\bibfnamefont{G.}~\bibnamefont{Biroli}},
  \bibinfo{editor}{\bibfnamefont{J.-P.} \bibnamefont{Bouchaud}},
  \bibinfo{editor}{\bibfnamefont{L.}~\bibnamefont{Cipelletti}},
  \bibnamefont{and} \bibinfo{editor}{\bibfnamefont{W.}~\bibnamefont{van
  Saarloos}} (\bibinfo{publisher}{Oxford University Press},
  \bibinfo{year}{2011}).

\bibitem[{\citenamefont{Goldstein}(1969)}]{Go69}
\bibinfo{author}{\bibfnamefont{M.}~\bibnamefont{Goldstein}},
  \bibinfo{journal}{The Journal of Chemical Physics}
  \textbf{\bibinfo{volume}{51}}, \bibinfo{pages}{3728} (\bibinfo{year}{1969}),
  \urlprefix\url{http://link.aip.org/link/?JCP/51/3728/1}.

\bibitem[{\citenamefont{Stillinger and Weber}(1982)}]{SW82}
\bibinfo{author}{\bibfnamefont{F.~H.} \bibnamefont{Stillinger}}
  \bibnamefont{and} \bibinfo{author}{\bibfnamefont{T.~A.} \bibnamefont{Weber}},
  \bibinfo{journal}{Phys. Rev. A} \textbf{\bibinfo{volume}{25}},
  \bibinfo{pages}{978} (\bibinfo{year}{1982}).

\bibitem[{\citenamefont{Biroli and Bouchaud}(2012)}]{BB09}
\bibinfo{author}{\bibfnamefont{G.}~\bibnamefont{Biroli}} \bibnamefont{and}
  \bibinfo{author}{\bibfnamefont{J.}~\bibnamefont{Bouchaud}}, in
  \emph{\bibinfo{booktitle}{Structural Glasses and Supercooled Liquids: Theory,
  Experiment and Applications}}, edited by
  \bibinfo{editor}{\bibnamefont{P.G.Wolynes}} \bibnamefont{and}
  \bibinfo{editor}{\bibnamefont{V.Lubchenko}} (\bibinfo{publisher}{Wiley \&
  Sons}, \bibinfo{year}{2012}), \eprint{{\tt arXiv:0912.2542}}.

\bibitem[{\citenamefont{Ritort and Sollich}(2003)}]{RS03}
\bibinfo{author}{\bibfnamefont{F.}~\bibnamefont{Ritort}} \bibnamefont{and}
  \bibinfo{author}{\bibfnamefont{P.}~\bibnamefont{Sollich}},
  \bibinfo{journal}{Advances in Physics} \textbf{\bibinfo{volume}{52}},
  \bibinfo{pages}{219} (\bibinfo{year}{2003}).

\bibitem[{\citenamefont{Biroli and Garrahan}(2013)}]{BG13}
\bibinfo{author}{\bibfnamefont{G.}~\bibnamefont{Biroli}} \bibnamefont{and}
  \bibinfo{author}{\bibfnamefont{J.~P.} \bibnamefont{Garrahan}},
  \bibinfo{journal}{The Journal of chemical physics}
  \textbf{\bibinfo{volume}{138}}, \bibinfo{pages}{12A301}
  (\bibinfo{year}{2013}).

\bibitem[{\citenamefont{G{\"o}tze}(2009)}]{Go09}
\bibinfo{author}{\bibfnamefont{W.}~\bibnamefont{G{\"o}tze}},
  \emph{\bibinfo{title}{Complex dynamics of glass-forming liquids: A
  mode-coupling theory}}, vol. \bibinfo{volume}{143}
  (\bibinfo{publisher}{Oxford University Press, USA}, \bibinfo{year}{2009}).

\bibitem[{\citenamefont{M{\'e}zard and Parisi}(1996)}]{MP96}
\bibinfo{author}{\bibfnamefont{M.}~\bibnamefont{M{\'e}zard}} \bibnamefont{and}
  \bibinfo{author}{\bibfnamefont{G.}~\bibnamefont{Parisi}},
  \bibinfo{journal}{Journal of Physics A: Mathematical and General}
  \textbf{\bibinfo{volume}{29}}, \bibinfo{pages}{6515} (\bibinfo{year}{1996}).

\bibitem[{\citenamefont{Parisi and Zamponi}(2010)}]{PZ10}
\bibinfo{author}{\bibfnamefont{G.}~\bibnamefont{Parisi}} \bibnamefont{and}
  \bibinfo{author}{\bibfnamefont{F.}~\bibnamefont{Zamponi}},
  \bibinfo{journal}{Rev. Mod. Phys.} \textbf{\bibinfo{volume}{82}},
  \bibinfo{pages}{789} (\bibinfo{year}{2010}).

\bibitem[{\citenamefont{Szamel}(2010)}]{Sz10}
\bibinfo{author}{\bibfnamefont{G.}~\bibnamefont{Szamel}}, \bibinfo{journal}{EPL
  (Europhysics Letters)} \textbf{\bibinfo{volume}{91}}, \bibinfo{pages}{56004}
  (\bibinfo{year}{2010}).

\bibitem[{\citenamefont{Caltagirone et~al.}(2012)\citenamefont{Caltagirone,
  Ferrari, Leuzzi, Parisi, Ricci-Tersenghi, and Rizzo}}]{CFLPRR12}
\bibinfo{author}{\bibfnamefont{F.}~\bibnamefont{Caltagirone}},
  \bibinfo{author}{\bibfnamefont{U.}~\bibnamefont{Ferrari}},
  \bibinfo{author}{\bibfnamefont{L.}~\bibnamefont{Leuzzi}},
  \bibinfo{author}{\bibfnamefont{G.}~\bibnamefont{Parisi}},
  \bibinfo{author}{\bibfnamefont{F.}~\bibnamefont{Ricci-Tersenghi}},
  \bibnamefont{and} \bibinfo{author}{\bibfnamefont{T.}~\bibnamefont{Rizzo}},
  \bibinfo{journal}{Phys. Rev. Lett.} \textbf{\bibinfo{volume}{108}},
  \bibinfo{pages}{085702} (\bibinfo{year}{2012}).

\bibitem[{\citenamefont{Franz et~al.}(2012)\citenamefont{Franz, Jacquin,
  Parisi, Urbani, and Zamponi}}]{FJPUZ12}
\bibinfo{author}{\bibfnamefont{S.}~\bibnamefont{Franz}},
  \bibinfo{author}{\bibfnamefont{H.}~\bibnamefont{Jacquin}},
  \bibinfo{author}{\bibfnamefont{G.}~\bibnamefont{Parisi}},
  \bibinfo{author}{\bibfnamefont{P.}~\bibnamefont{Urbani}}, \bibnamefont{and}
  \bibinfo{author}{\bibfnamefont{F.}~\bibnamefont{Zamponi}},
  \bibinfo{journal}{Proceedings of the National Academy of Sciences}
  \textbf{\bibinfo{volume}{109}}, \bibinfo{pages}{18725}
  (\bibinfo{year}{2012}).

\bibitem[{\citenamefont{Franz et~al.}(2013)\citenamefont{Franz, Jacquin,
  Parisi, Urbani, and Zamponi}}]{FJPUZ12Long}
\bibinfo{author}{\bibfnamefont{S.}~\bibnamefont{Franz}},
  \bibinfo{author}{\bibfnamefont{H.}~\bibnamefont{Jacquin}},
  \bibinfo{author}{\bibfnamefont{G.}~\bibnamefont{Parisi}},
  \bibinfo{author}{\bibfnamefont{P.}~\bibnamefont{Urbani}}, \bibnamefont{and}
  \bibinfo{author}{\bibfnamefont{F.}~\bibnamefont{Zamponi}},
  \bibinfo{journal}{The Journal of chemical physics}
  \textbf{\bibinfo{volume}{138}}, \bibinfo{pages}{12A540}
  (\bibinfo{year}{2013}).

\bibitem[{\citenamefont{Kurchan et~al.}(2013)\citenamefont{Kurchan, Parisi,
  Urbani, and Zamponi}}]{KPUZ13}
\bibinfo{author}{\bibfnamefont{J.}~\bibnamefont{Kurchan}},
  \bibinfo{author}{\bibfnamefont{G.}~\bibnamefont{Parisi}},
  \bibinfo{author}{\bibfnamefont{P.}~\bibnamefont{Urbani}}, \bibnamefont{and}
  \bibinfo{author}{\bibfnamefont{F.}~\bibnamefont{Zamponi}},
  \bibinfo{journal}{The Journal of Physical Chemistry B}
  (\bibinfo{year}{2013}).

\bibitem[{\citenamefont{Cugliandolo et~al.}(1997)\citenamefont{Cugliandolo,
  Kurchan, and Peliti}}]{CKP97}
\bibinfo{author}{\bibfnamefont{L.}~\bibnamefont{Cugliandolo}},
  \bibinfo{author}{\bibfnamefont{J.}~\bibnamefont{Kurchan}}, \bibnamefont{and}
  \bibinfo{author}{\bibfnamefont{L.}~\bibnamefont{Peliti}},
  \bibinfo{journal}{Physical Review E} \textbf{\bibinfo{volume}{55}},
  \bibinfo{pages}{3898} (\bibinfo{year}{1997}).

\bibitem[{\citenamefont{Franz and Virasoro}(2000)}]{FV00}
\bibinfo{author}{\bibfnamefont{S.}~\bibnamefont{Franz}} \bibnamefont{and}
  \bibinfo{author}{\bibfnamefont{M.}~\bibnamefont{Virasoro}},
  \bibinfo{journal}{Journal of Physics A: Mathematical and General}
  \textbf{\bibinfo{volume}{33}}, \bibinfo{pages}{891} (\bibinfo{year}{2000}).

\bibitem[{\citenamefont{Franz and Parisi}(2013)}]{FP12}
\bibinfo{author}{\bibfnamefont{S.}~\bibnamefont{Franz}} \bibnamefont{and}
  \bibinfo{author}{\bibfnamefont{G.}~\bibnamefont{Parisi}},
  \bibinfo{journal}{Journal of Statistical Mechanics: Theory and Experiment}
  \textbf{\bibinfo{volume}{2013}}, \bibinfo{pages}{P02003}
  (\bibinfo{year}{2013}).

\bibitem[{\citenamefont{Krzakala and Kurchan}(2007)}]{KK07}
\bibinfo{author}{\bibfnamefont{F.}~\bibnamefont{Krzakala}} \bibnamefont{and}
  \bibinfo{author}{\bibfnamefont{J.}~\bibnamefont{Kurchan}},
  \bibinfo{journal}{Physical Review E (Statistical, Nonlinear, and Soft Matter
  Physics)} \textbf{\bibinfo{volume}{76}}, \bibinfo{eid}{021122}
  (pages~\bibinfo{numpages}{13}) (\bibinfo{year}{2007}),
  \urlprefix\url{http://link.aps.org/abstract/PRE/v76/e021122}.

\bibitem[{\citenamefont{Krzakala and Zdeborova}(2011)}]{KZ11}
\bibinfo{author}{\bibfnamefont{F.}~\bibnamefont{Krzakala}} \bibnamefont{and}
  \bibinfo{author}{\bibfnamefont{L.}~\bibnamefont{Zdeborova}},
  \bibinfo{journal}{The Journal of Chemical Physics}
  \textbf{\bibinfo{volume}{134}}, \bibinfo{eid}{034513}
  (pages~\bibinfo{numpages}{13}) (\bibinfo{year}{2011}),
  \urlprefix\url{http://link.aip.org/link/?JCP/134/034513/1}.

\bibitem[{\citenamefont{Cugliandolo and Kurchan}(1993)}]{CK93}
\bibinfo{author}{\bibfnamefont{L.~F.} \bibnamefont{Cugliandolo}}
  \bibnamefont{and} \bibinfo{author}{\bibfnamefont{J.}~\bibnamefont{Kurchan}},
  \bibinfo{journal}{Phys. Rev. Lett.} \textbf{\bibinfo{volume}{71}},
  \bibinfo{pages}{173} (\bibinfo{year}{1993}).

\bibitem[{\citenamefont{G{\"o}tze et~al.}(1991)\citenamefont{G{\"o}tze, Hansen,
  Levesque, and Zinn-Justin}}]{gotze1991liquids}
\bibinfo{author}{\bibfnamefont{W.}~\bibnamefont{G{\"o}tze}},
  \bibinfo{author}{\bibfnamefont{J.}~\bibnamefont{Hansen}},
  \bibinfo{author}{\bibfnamefont{D.}~\bibnamefont{Levesque}}, \bibnamefont{and}
  \bibinfo{author}{\bibfnamefont{J.}~\bibnamefont{Zinn-Justin}},
  \emph{\bibinfo{title}{Liquids, freezing and the glass transition}}
  (\bibinfo{year}{1991}).

\bibitem[{\citenamefont{Szamel and L{\"o}wen}(1991)}]{szamel1991mode}
\bibinfo{author}{\bibfnamefont{G.}~\bibnamefont{Szamel}} \bibnamefont{and}
  \bibinfo{author}{\bibfnamefont{H.}~\bibnamefont{L{\"o}wen}},
  \bibinfo{journal}{Physical Review A} \textbf{\bibinfo{volume}{44}},
  \bibinfo{pages}{8215} (\bibinfo{year}{1991}).

\bibitem[{\citenamefont{Given and Stell}(1993)}]{GS93}
\bibinfo{author}{\bibfnamefont{J.~A.} \bibnamefont{Given}} \bibnamefont{and}
  \bibinfo{author}{\bibfnamefont{G.}~\bibnamefont{Stell}},
  \bibinfo{journal}{Condensed Matter Theories} \textbf{\bibinfo{volume}{8}},
  \bibinfo{pages}{395} (\bibinfo{year}{1993}).

\bibitem[{\citenamefont{Given}(1992)}]{Gi92}
\bibinfo{author}{\bibfnamefont{J.~A.} \bibnamefont{Given}},
  \bibinfo{journal}{Phys. Rev. A} \textbf{\bibinfo{volume}{45}},
  \bibinfo{pages}{816} (\bibinfo{year}{1992}),
  \urlprefix\url{http://link.aps.org/doi/10.1103/PhysRevA.45.816}.

\bibitem[{\citenamefont{Hansen and McDonald}(1986)}]{Hansen}
\bibinfo{author}{\bibfnamefont{J.-P.} \bibnamefont{Hansen}} \bibnamefont{and}
  \bibinfo{author}{\bibfnamefont{I.~R.} \bibnamefont{McDonald}},
  \emph{\bibinfo{title}{Theory of simple liquids}}
  (\bibinfo{publisher}{Academic Press}, \bibinfo{address}{London},
  \bibinfo{year}{1986}).

\bibitem[{\citenamefont{Cugliandolo}(2003)}]{Cu03}
\bibinfo{author}{\bibfnamefont{L.}~\bibnamefont{Cugliandolo}},
  \bibinfo{journal}{Les Houches-\'Ecole d' \'Et\' e de Physique Th\'eorique}
  \textbf{\bibinfo{volume}{77}}, \bibinfo{pages}{367} (\bibinfo{year}{2003}).

\bibitem[{\citenamefont{Rizzo}(2013)}]{Ri13}
\bibinfo{author}{\bibfnamefont{T.}~\bibnamefont{Rizzo}},
  \bibinfo{journal}{Physical review. E, Statistical, nonlinear, and soft matter
  physics} \textbf{\bibinfo{volume}{87}}, \bibinfo{pages}{022135}
  (\bibinfo{year}{2013}).

\bibitem[{\citenamefont{Latz}(2000)}]{La00}
\bibinfo{author}{\bibfnamefont{A.}~\bibnamefont{Latz}},
  \bibinfo{journal}{Journal of Physics: Condensed Matter}
  \textbf{\bibinfo{volume}{12}}, \bibinfo{pages}{6353} (\bibinfo{year}{2000}).

\bibitem[{\citenamefont{Kurchan et~al.}(2012)\citenamefont{Kurchan, Parisi, and
  Zamponi}}]{KPZ12}
\bibinfo{author}{\bibfnamefont{J.}~\bibnamefont{Kurchan}},
  \bibinfo{author}{\bibfnamefont{G.}~\bibnamefont{Parisi}}, \bibnamefont{and}
  \bibinfo{author}{\bibfnamefont{F.}~\bibnamefont{Zamponi}},
  \bibinfo{journal}{J. Stat. Mech. P10012}  (\bibinfo{year}{2012}).

\bibitem[{\citenamefont{Franz et~al.}(2011)\citenamefont{Franz, Parisi,
  Ricci-Tersenghi, and Rizzo}}]{FPRR11}
\bibinfo{author}{\bibfnamefont{S.}~\bibnamefont{Franz}},
  \bibinfo{author}{\bibfnamefont{G.}~\bibnamefont{Parisi}},
  \bibinfo{author}{\bibfnamefont{F.}~\bibnamefont{Ricci-Tersenghi}},
  \bibnamefont{and} \bibinfo{author}{\bibfnamefont{T.}~\bibnamefont{Rizzo}},
  \bibinfo{journal}{The European Physical Journal E: Soft Matter and Biological
  Physics} \textbf{\bibinfo{volume}{34}}, \bibinfo{pages}{1}
  (\bibinfo{year}{2011}).

\end{thebibliography}

\end{document}